\documentstyle[aps,prl,floats,graphicx,epsfig]{revtex}

\begin{document}
\draft
\title{Simulation of Panlev\'e-Gullstrand black hole in thin $^3$He-A film.}
\author{G.E. Volovik\\
Low Temperature Laboratory, Helsinki University of
Technology,
P.O.Box 2200, FIN-02015 HUT, Finland\\
L.D. Landau Institute for
Theoretical Physics,  Kosygin Str. 2, 117940 Moscow, Russia
}
\maketitle
\begin{abstract}
{The quasi-stationary superfluid state is constructed, which exhibits the
event horizon and Hawking radiation.
 }
\end{abstract}

\tableofcontents

\subsection{\it  Introduction}

It is well known that the gravitational field can be simulated in condensed
matter by the motion of the liquid:  propagation of some perturbations in the
moving liquid obeys the same equation as propagation of relativistic particles
in the gravitational field. These
perturbations are sound waves in normal fluids \cite{UnruhSonic,Visser1999}
and quasiparticles in superfluids \cite{JacobsonVolovik} (phonons in
superfluid
$^4$He and low-energy
Bogoliubov fermions in superfluid
$^3$He-A). If the fluid motion is radial and spherically symmetric the
effective metric is expressed in terms of the radial velocity $v(r)$ as:
\begin{equation}
 ds^2=-\left(c^2- v^2(r)\right)dt^2+ 2v(r)drdt+dr^2+r^2d\Omega^2~.
\label{SphericallySymmetricField}
\end{equation}
For superfluids  $v$ is the velocity of superfluid vacuum $v_s$.

The kinetic energy of superflow plays the part of the gravitational
potential: $ \Phi=- v^2(r)/2$. If one chooses the velocity field corresponding
to the potential of the point body of mass $M$
\begin{equation}
v^2(r) = -2\Phi={2GM \over r}\equiv c^2{r_h\over r}~,
\label{Schwarzschild}
\end{equation}
one obtains the Panlev\'e-Gullstrand form of
Schwarzschild geometry (see e.g.
ref.\cite{Visser1999}).
Here $r_h$ denotes the position of the event horizon, where the velocity
reaches the "speed of light"  ($c$ is the speed of sound for phonons or the
slope in the energy spectrum $E=\pm cp$ of Bogoliubov fermions). If the fluid
moves towards the origin, i.e. $v(r)<0$, this velocity field reproduces the
horizon of the black hole (the so called sonic black hole \cite{UnruhSonic}):
Since the velocity of the fluid behind horizon exceeds the speed $c$ of the
propagation of the perturbations with respect to the fluid, the low-energy
quasipartilcles are trapped within the horizon. In quantum fermi liquids --
superfluid phases of
$^3$He -- such kind of the hydrodynamic black hole will allow us to investigate
the quantum fermionic vacuum in a classical gravitational field in the presence
of a horizon.

The hydrodynamic black hole was first suggested by Unruh for ordinary liquid
\cite{UnruhSonic}. However since all the known  normal liquids are
classical, the
most interesting quantum effects related to the horizon cannot be simulated in
such flow. Also the geometry is such that it cannot be realized:  in the radial
flow inward the liquid becomes accumulated at the origin, that is why this
sonic
black hole cannot be stationary.  In the other scenario a horizon appears in
moving solitons, if the velocity of the soliton exceeds the local "speed of
light"
\cite{JacobsonVolovik}. This scenario has the same drawback: in finite
system the
motion of the soliton cannot be supported for a long time.  In a draining
bathtub
geometry suggested in Ref.\cite{Visser1997} the fluid motion can be made
constant
in time. However the friction of the liquid, which moves through the drain, is
the main source of dissipation. This will hide any quantum effects related
to the
horizon. The superfluidity of the liquid does not help much in this situation:
Horizon does not appear: The "superluminal" (supercritical) motion with
respect to the boundaries is unstable, because the interaction with the walls
produces the Cherenkov radiation of quasiparticles, and superfluidity collapses
(see \cite{KopninVolovik}).

Here we suggest a scenario, in which this collapse is
avoided. The superfluid motion becomes quasi-stationary and exhibits the event
horizon; the life time of the "superluminally" flowing state is determined  by
intrinsic mechanisms related to existence of a horizon, in particular by the
analogue of Hawking radiation.

\subsection{\it  Simulation of 2D black hole}

\begin{figure}[!!!t]
%\centerline{\epsfxsize=0.90\textwidth\epsfbox{2DBHJETPL.eps}}
%\bigskip
\begin{center}
\leavevmode
\epsfig{file=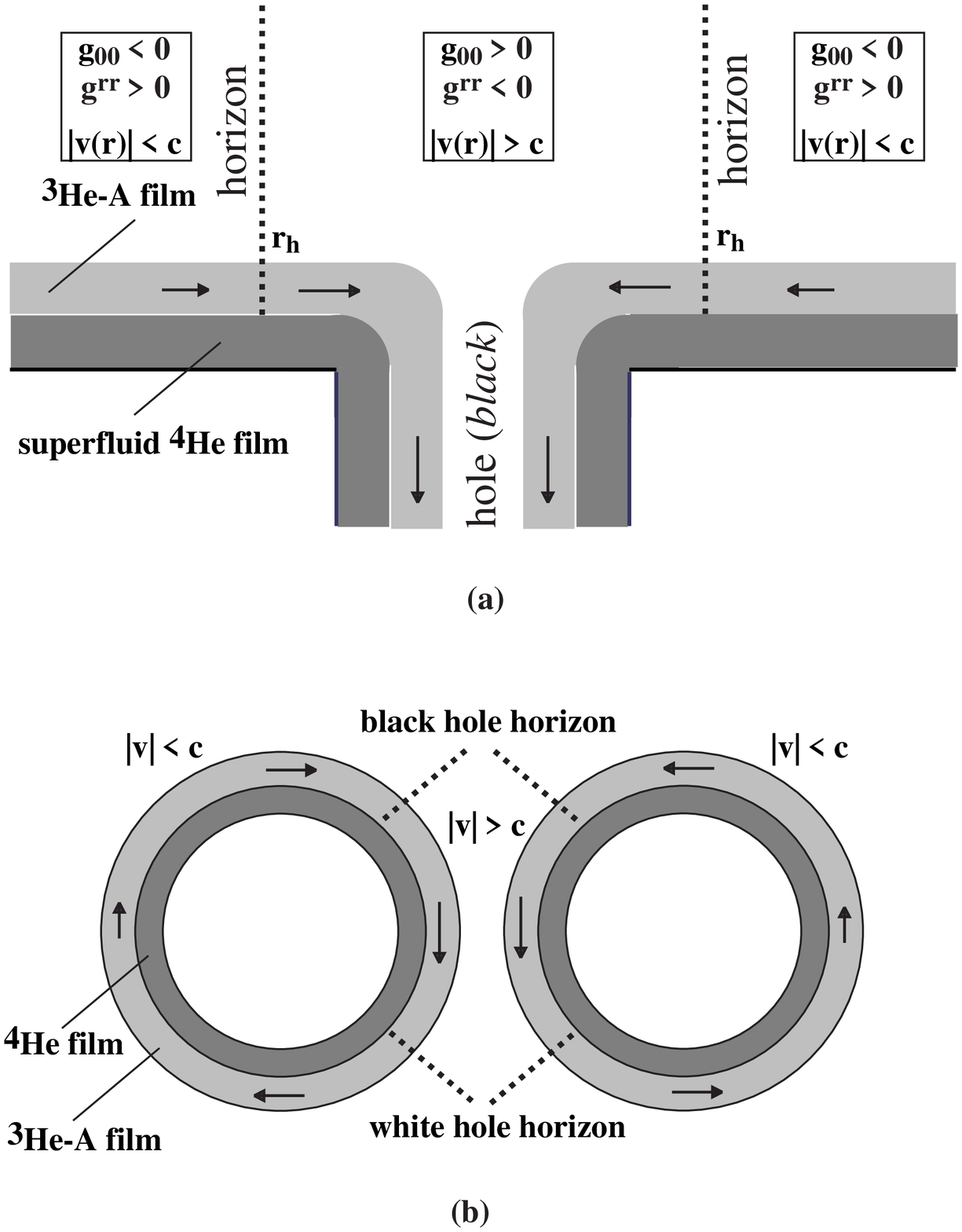,width=0.8\linewidth}
\caption[2DBH]
    {Similation of 2D black hole in thin $^3$He-A film. (a) Draining bathtub
geometry. (b) $^3$He-A film circulating on the top of the $^4$He film on
torus. }
\label{2DBH}
\end{center}
\end{figure}

 The stationary black hole can be realized
in the following geometry, which is the development of the  bathtub geometry of
Ref.\cite{Visser1997} (see Fig.
\ref{2DBH}(a)).  The superfluid
$^3$He-A film is moving towards the center of the disk (i.e. $v(r)<0$),
where it
escapes to the third dimension due to the orifice (hole). If the thickness
of the film is constant, the flow velocity of the 2D motion increases
towards the
center as
$v(r)=a/r$ and at $r=r_h=a/c$ it reaches the "speed of light" $c$  (now $r$
denotes the radial coordinate in the cylindrical system). If  this happens the
hole becomes the black hole:  Behind the horizon, at $r<r_h$,  the
(quasi)particles can move only to the hole (orifice), since their velocity of
propagation with respect to the superfluid condensate is less than the velocity
$v$ of the condensate.

The black hole analogy is also supported by the effective metric experienced by
the quasiparticles.  The energy spectrum of the low-energy Bogoliubov
fermions is
given by
\begin{equation}
(E-{\bf p}\cdot{\bf v})^2 =c^2(p_x^2+p_y^2) + v_F^2(p_z\mp p_F)^2~.
\label{rel}
\end{equation}
Here the axis $z$ is along the normal of the film. This axis $z$ also marks the
direction of the unit orbital vector
$\hat l$, which is the anisotropy axis for the "speed of light": $\hat l$ is
fixed along the normal to the film. The "speed of light" for
quasiparticles propagating along the film is $c \sim 3~$cm/sec. It is much
smaller than the Fermi velocity
$v_F$ which corresponds to the "speed of light" for   quasiparticles
propagating
along the normal to the film. This $c$ is also much smaller than the speed of
sound in
$^3$He-A, that is why the motion of fluid has no effect on the density of the
liquid.

Outside the orifice the velocity  of the superfluid ${\bf v}$ is two
dimensional and radial. With such velocity field the energy spectrum in
Eq.(\ref{rel}) corresponds to the motion of the Bogoliubov quasiparticle in the
space with following effective metric
\begin{equation}
 ds^2=-\left(c^2- v^2(r)\right)dt^2+ 2v(r)drdt+
dr^2+r^2d\phi^2 +{c^2\over v_F^2}dz^2~.
\label{CylindricallySymmetricField}
\end{equation}
Across the horizon the $g_{00}$ component of the metric changes sign, which
marks the presence of the horizon at $r=r_h$, where $v(r_h)=c$.

%%%%%%%%%%%%%%%%%%%%%%%%%%%%%%%%%%%%%%%%%%%%%%%%%%%%%%%%%%%
%%%%%%%%%%%%%%%%%%%%%%%%%%%%%%%%%%%%%%%%%%%%%%%%%%%%%%%%%%

The important element of the construction in Fig.
\ref{2DBH}(a) is that the moving superfluid  $^3$He-A film
is placed on the top of the superfluid $^4$He film. This is made to avoid the
interaction of the $^3$He-A film with the solid substrate.  The superfluid
$^4$He
film effectively screens the interaction and thus prevents the collapse of the
"superluminal" flow of $^3$He-A.

The motion of the superfluid $^3$He-A with respect to superfluid $^4$He film is
not dangerous: The superfluid $^4$He is not excited even if $^3$He-A moves with
its superluminal velocity: $c$ for $^3$He-A is much smaller than the Landau
velocity for radiation of quasiparticles in superfluid $^4$He, which is about
$50$ m/sec. In this consideration we neglected the radiation of surface waves,
assuming that the thickness of $^4$He film is small enough.

Finally one can close the superflow by introducing the toroidal geometry.  Fig.
\ref{2DBH}(b) shows the superflow around meridians (minor circles) of the torus
in the cross-sectional plane. Both superfluid condensates, $^4$He and $^3$He-A,
circulate around meridians with integer numbers
$N_4$ and $N_3$ of superfluid velocity circulation quanta,
$\kappa_4=2\pi\hbar/m_4$ and $\kappa_3= \pi\hbar/m_3$.
If the inner radius of the torus is small, the superfluid velocity is enhanced
in the region close to the inner circle, where it can exceed $c$. In this case
both the black hole horizon and the white hole horizon appear.

Since the  extrinsic
mechanism of the friction of $^3$He-A film -- the  scattering of
quasiparticles on
the roughness of substrate -- is abandoned, we can consider now intrinsic
mechanisms of dissipation of the supecritical flow. The most interesting one is
the Hawking radiation related to existence of a horizon.

\subsection{\it Vacuum in comoving and rest frames.}

Let us consider the simplest case of the 2D motion along the film in the
bathtub
geometry of Fig. \ref{2DBH}(a). This can be easily generalized to the motion in
the torus geometry.

\begin{figure}[!!!t]
%\centerline{\epsfxsize=0.70\textwidth\epsfbox{VacuumAcrossHor.eps}}
%\bigskip
\begin{center}
\leavevmode
\epsfig{file=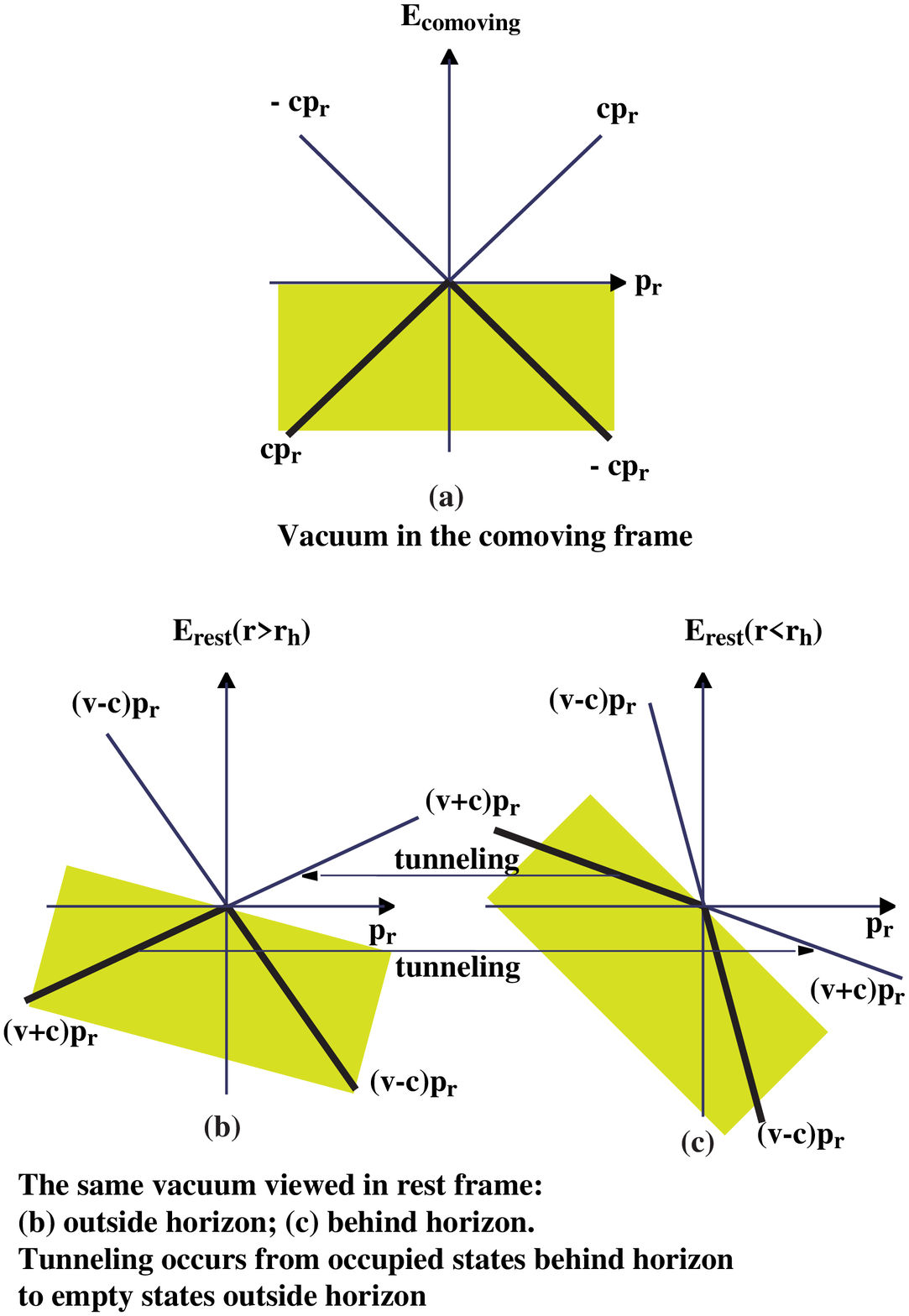,width=0.7\linewidth}
\caption[Vacua]
    {(a) Fermionic vacuum in the comoving frame. The states with $E_{\rm
com}<0$
are occupied (thick lines). The same vacuum viewed in the rest frame (b)
outside
horizon and (c) inside horizon. Behind the horizon the branch
$E_{\rm rest}=(v+c)p_r$ (for $p_\perp =0$) has inverse population as seen
in the
rest frame: the states with positive
enegry
$E_{\rm rest}>0$ are filled, while the states with
$E_{\rm rest}<0$ are empty. The tunneling across horizon from the occupied
states
to the empty states with the same energy gives rise to the Hawking
radiation from the horizon.}
\label{Vacua}
\end{center}
\end{figure}

There are two important reference frames:  (i) The
frame of the  observer, who is locally comoving with the superfluid vacuum.
In this frame the local superfluid velocity is zero,  $v=0$, so that the energy
spectrum of the  Bogoliubov fermions in the place of the observer is (here
we assume a pure 2D motion along the film)
\begin{equation}
E_{\rm com}=\pm cp~.
\label{EnergyComovingFrame}
\end{equation}
In this geometry, in which the superflow velocity is confined in the plane
of the
film, the speed $c$ coincides with the Landau critical velocity of the
superfluid vacuum, $v_{\rm Landau}={\rm min} (|E_{\rm com}(p)|/p)$. The
vacuum as
determined by the comoving observer is shown on Fig.~\ref{Vacua}(a): fermions
occupy the negative energy levels in the Dirac sea (the states with the minus
sign in Eq.(\ref{EnergyComovingFrame}). It is the counterpart of the Minkowski
vacuum, which is however determined only locally: The comoving frame cannot be
determined globally. Moreover for the comowing observer, whose velocity changes
with time, the whole velocity field
${\bf v}({\bf r},t)$ of the superflow is time dependent. This does not allow to
determine the energy correctly.

(ii) The energy can be well defined in the laboratory frame (the
rest frame). In this frame the system is stationary, though is not static: The
effective metric does not depend on time,  so that the energy is conserved,
but this metric contains the mixed component
$g_{0i}$. The energy in the rest frame is
obtained from the local energy in the comoving frame by the Doppler shift. In
case of the radial superflow one has
\begin{equation}
E_{\rm rest}=\pm cp + p_rv(r)~.
\label{Energy}
\end{equation}

Figs.~\ref{Vacua}(b-c) show how the "Minkowski" vacuum of the comoving frame is
seen by the rest observer (note that the velocity is negative, $v(r)<0$).
In the absense of horizon, or outside the horizon the local
vacuum does not change: the states which are occupied (empty) in the Minkowski
vacuum remain occupied (empty) in the rest frame vacuum (see
Fig.~\ref{Vacua}(b)).  In the presense of horizon behind which the velocity of
superflow exceeds the Landau critical velocity the situation changes:
Behind the
horizon the vacuum in the rest frame differs from that in the comoving
frame. Let
us for simplicity consider the states with zero transverse momentum
$p_\phi=0$ on the branch $E_{\rm rest}=(v(r)+c)p_r$ in the rest frame. If the
system is in  the Minkowski vacuum state (i.e. in the ground state as viewed by
comoving observer),  quasiparticles on this branch has reversed distribution in
the rest frame: the negative energy states are empty, while the  positive
energy
states are occupied  (see  Fig.~\ref{Vacua}(c)). For this branch the particle
distribution corresponds to the negative temperature
$T=-0$ behind horizon.

Since the energy in the rest frame is a good quantum number, the fermions
can tunnel across the horizon from the occupied levels to the empty ones
with the
same energy. Thus if the system is initially in the Minkowski vacuum in the
comoving frame, the tunneling disturbs this vacuum state: Pairs of excitations
are created: the  quasiparticle, say, is created outside the horizon while its
partner -- the quasihole -- is created inside the horizon. This simulates the
Hawking radiation from the black hole.

%%%%%%%%%%%%%%%%%%%%%%%%%%%%%%%%%%%%%%%%%%%%%%%%%%%%%%%%%%%
%%%%%%%%%%%%%%%%%%%%%%%%%%%%%%%%%%%%%%%%%%%%%%%%%%%%%%%%%%

\subsection{\it  Hawking radiation}

%%%%%%%%%%%%%%%%%%%%%%%%%%%%%%%%%%%%%%%%%%%%%%%%%%%%%%%%%%%
\begin{figure}[!!!t]
%\centerline{\epsfxsize=0.90\textwidth\epsfbox{TunnelingTrajJETPL.eps}}
%\bigskip
\begin{center}
\leavevmode
\epsfig{file=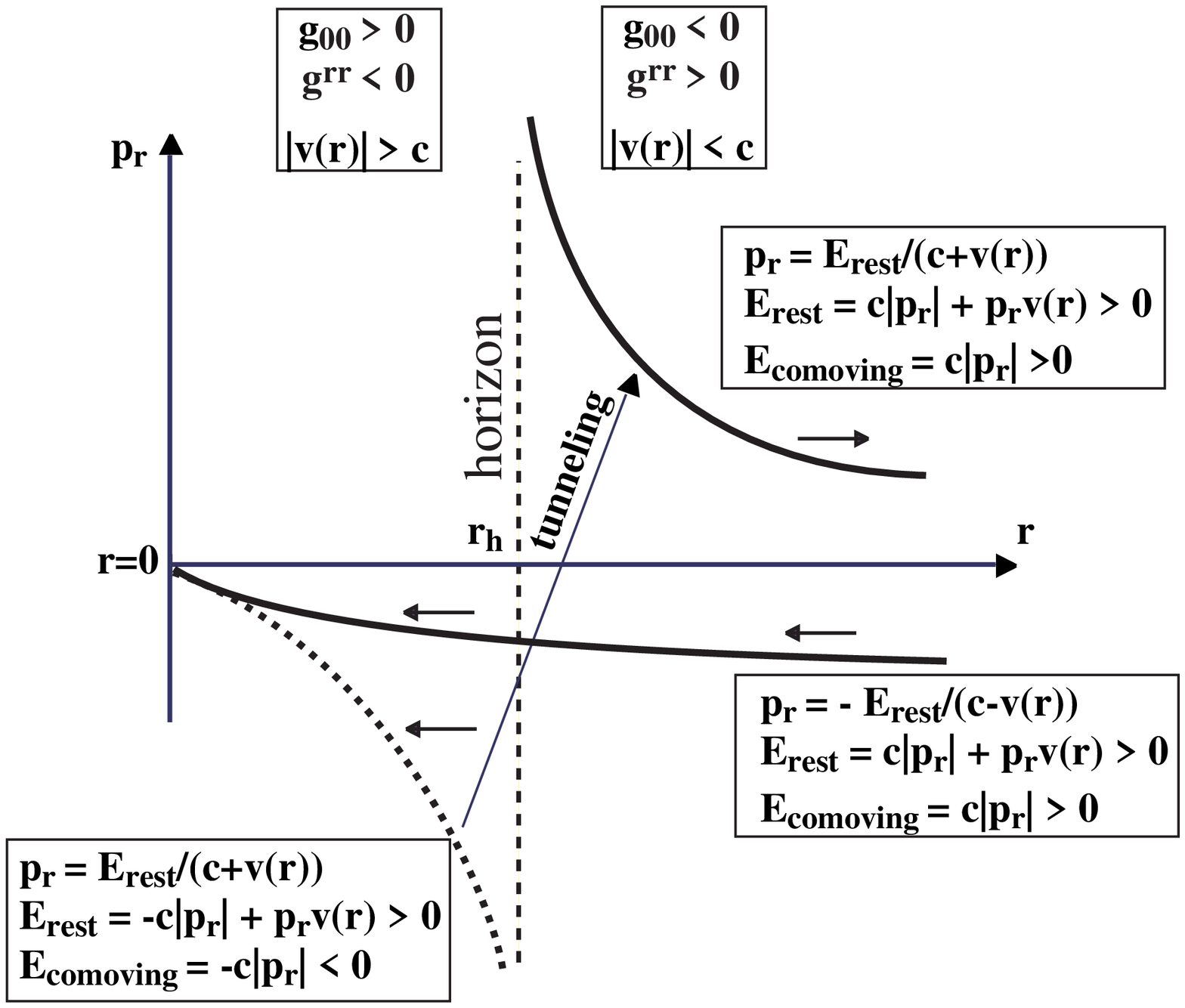,width=0.8\linewidth}
\caption[TunnelingTraj]
    {Tunneling from Minkowski vacuum within the horizon to the outgoing
mode. }
\label{TunnelingTraj}
\end{center}
\end{figure}
%%%%%%%%%%%%%%%%%%%%%%%%%%%%%%%%%%%%%%%%%%%%%%%%%%%%%%%%%%

To estimate the tunneling rate in the semiclassical approximation, let us
consider
the classical trajectories $p_r(r)$ of particles, say, with positive energy,
$E_{\rm rest}>0$, for the simplest case when the transverse momentum
$p_\phi$ is
zero Fig.~\ref{TunnelingTraj}. The branch $E_{\rm rest}=(v(r)-c)p_r$ describes
the  incoming particles with
$p_r<0$ which propagate through the horizon to the orifice (or to the
singularity at $r=0$, if the orifice is infinitely small) without any
singularity
at the horizon. The classical trajectories of these particles are
\begin{equation}
p_r(r)=-{E_{\rm rest}\over c-v(r)} <0 ~.
\label{IncomingTrajectory}
\end{equation}
The energies of these particles viewed by the comoving observer are also
positive:
$E_{\rm com}(r)= -cp_r(r) = E_{\rm rest}(c/c-v(r)) >0$.

Another branch  $E_{\rm rest}=(v(r)+c)p_r$ in Fig.~\ref{TunnelingTraj} contains
two disconnected pieces describing the particle propagating from the horizon in
two opposite directions:
\begin{eqnarray}
r>r_h~:~p_r(r)= {E_{\rm rest}\over c+v(r)}>0 ~, ~ E_{\rm com}(r)= cp_r(r)>0~.
\label{TunnelingTrajectory2}
\\
r<r_h~:~p_r(r)= {E_{\rm rest}\over c+v(r)}<0 ~,  ~E_{\rm
com}(r)= c p_r(r)<0~,
\label{TunnelingTrajectory1}
\end{eqnarray}
 The Eq.(\ref{TunnelingTrajectory2}) describes  the
outgoing particles -- the particles propagating from the horizon to the
exterior. The energy of these particles is positive in both frames, comoving
and rest.  The Eq.(\ref{TunnelingTrajectory1}) describes the propagation
of particles from the horizon to the orifice  (or to the singularity). Though
for the rest frame observer the energy of these particles is positive, these
particles, which live  within the horizon, belong to the Minkowski vacuum
in the
comoving frame.

The classical trajectory in
Eqs.(\ref{TunnelingTrajectory2},\ref{TunnelingTrajectory1}) is thus
disrupted at
the horizon. There is however a quantum mechanical transition between the
two pieces of the branch: the quantum tunneling. The tunneling amplitude can be
found in semiclassical approximation by shifting  the contour of integration to
the complex plane:
\begin{eqnarray}
w\sim  \exp (-2 S)~,\label{TunnelingProbability}\\
S={\bf Im}\int dr~ p_r(r)=  {\pi E_{\rm rest}\over | v'(r)|_{r=r_h}}~.
\label{TunnelingExponent}
\end{eqnarray}
This means that the wave function of any particle in the Minkowskii vacuum
inside the horizon contains an exponentially small part
describing the propagation from the horizon to infinity. This corresponds to
the radiation from the Minkowski vacuum in the presence of the event
horizon. The
exponential dependence  of the probability  on the quasiparticle energy
$E_{\rm rest}$ suggests that this radiation looks as thermal. The corresponding
temperature, the Hawking temperature, is
\begin{equation}
T_{\rm Hawking}= {\hbar | v'(r)|_{r=r_h} \over 2\pi}~.
\label{HawkingT}
\end{equation}

The radiation leads to the
quantum friction: the linear momentum of the flow decreases with time. This
occurs continuously until the superfluid Minkowski vacuum between the horizons
is completely exhausted and the superfluid state is violated. This leads to
the phase slip event, after which the number $N_3$ of circulation
quanta of superfluid velocity trapped by the torus is reduced. This process
will
repeatedly continue until the two horizons merge.

\subsection{\it  Negative temperature for the chiral 1+1 fermions.}

%%%%%%%%%%%%%%%%%%%%%%%%%%%%%%%%%%%%%%%%%%%%%%%%%%%%%%%%%%%
\begin{figure}[!!!t]
%\centerline{\epsfxsize=0.70\textwidth\epsfbox{VacuumRot.eps}}
%\bigskip
\begin{center}
\leavevmode
\epsfig{file=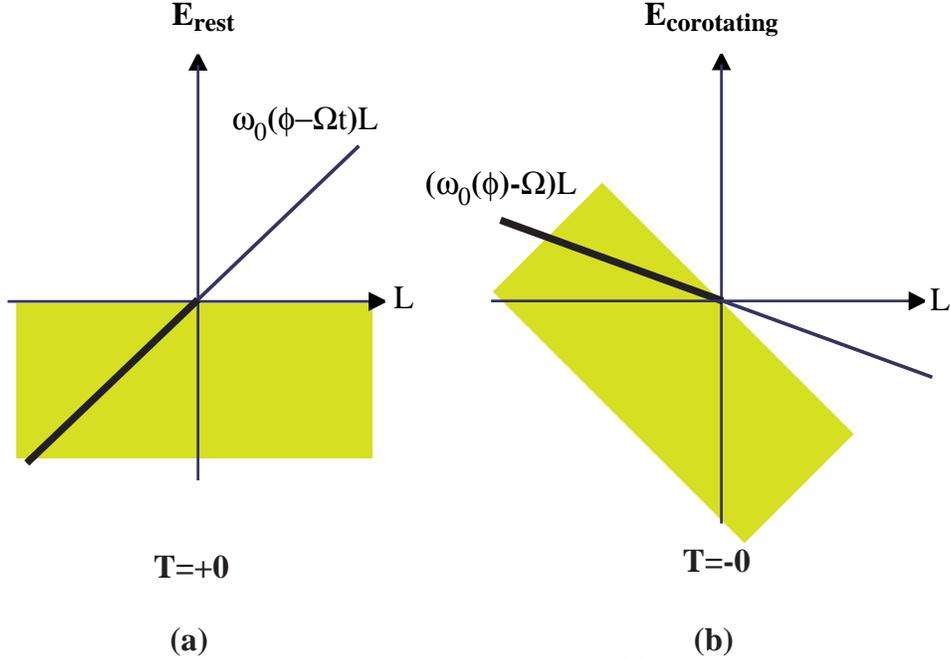,width=0.7\linewidth}
\caption[VacuumRot]
    {Enegry spectrum of fermionic quasiparticles in the vortex core. (a)
$\Omega=0$ corresponds to the nonrotating vortex. The spectrum is linear in the
angular momentum
$L$; $\omega_0(\phi)$ is the minigap, which depends on the angle $\phi$ if the
core is not axisymmetric. The states with negative $L$ are occupied. If the
vortex core is rotating with angular velocity
$\Omega$, then in the laboratory frame the minigap depends on
time. (b) The spectrum is well defined in the coordinate frame corotating with
the core. Here the spectrum is shown in the region of $\phi$ where
$\omega_0(\phi) <\Omega$, i.e. behind the horizon. If the initial state is the
Minkowski vacuum, it is seen by the corotating observer as the state with
$T=-0$.}
\label{VacuumRot}
\end{center}
\end{figure}
%%%%%%%%%%%%%%%%%%%%%%%%%%%%%%%%%%%%%%%%%%%%%%%%%%%%%%%%%%

It should be mentioned that there is an example where the negative temperature
behind a horizon is well defined. This is the case of the 1+1 dimensional
chiral
fermions living in the vortex core. For such fermions there is only one
branch,
$E=\omega_0(\phi)L$. Here $L$ is the angular momentum of the quasiparticle
in the
vortex core; $\omega_0(\phi)$ is the so-called minigap, which for the
nonaxisymmetric vortex core depends on  the azimuthal angle $\phi$. This is
equivalent to our branch
$E=c(r) p_r$ in the nonmoving liquid if the speed of light is coordinate
dependent. If the vortex core is rotated with  angular velocity $\Omega$, the
energy spectrum is time dependent in the laboratory frame
$E=\omega_0(\phi -\Omega t)L$. But it is time independent in the frame
corotating
with the vortex core, where the energy is well determined: $E_{\rm
corotating}=(\omega_0(\phi)  -\Omega)L$;   this is equivalent to our branch
$E_{\rm rest}=(v+c)p_r$ in the rest frame. The horizon can
occur  if the vortex core rotates with sufficently large angular velocity, such
that $\Omega$ exceeds the minimal value of the minigap
\cite{KopninVolovik2}. In this case  Since there is only one branch of the
fermions the negative temperature is well  determined. Behind the horizon the
Minkowski vacuum, which is the state with
$T=+0$ now in the laboratory frame  Fig.~\ref{VacuumRot}(a), is really the
state
with
$T=-0$ in the frame corotating with the core  Fig.~\ref{VacuumRot}(b); and vice
versa, the state with
$T=-0$ in the laboratory  frame is the state with $T=+0$ in
the corotating frame.

Such symmetry between the vacua for the 1+1 chiral
fermions suggests that there can be  also the symmetry between the nonzero
positive and negative temperatures.  Let us now take into account the Hawking
radiation and suppose that at infinity there is a heat bath with the
temperature
$T=T_{\rm Hawking}$. Then the heat flux from infinity exactly compensates
the radiation from the horizon. In such a metastable steady state the
distribution
of quasiparticles behind horizon would correspond to a nonzero negative
temperature $T=-T_{\rm Hawking}$.

\subsection{\it  Discussion.}

The above construction in Fig. \ref{2DBH}(b) allows us (at least in
principle) to
obtain the event horizon in the quasi-stationary regime, when the main
source of nonstationarity is the dissipation coming from the Hawking
radiation. As for the practical realization, there are, of course, many
technical problems to be solved. On the other hand, if the black hole
analog can in principle exist in condensed matter as the quasi-stationary
object,
its prototype -- the real black hole -- can also exist, at least in principle
(though it is not so easy to find the scenario of how this object can be
obtained
from the gravitational collapse of matter
\cite{Evanescent}).

The situation, in which the supercritical flow is described in terms
of the  event horizon and Hawking radiation, occurs only for the low-energy
fermions, whose spectrum is "relativistic".  Howeverever, the analogue of event
horizon persists even in the case of the "nonrelativistic" spectrum: the
"horizon" occurs at the surface where the local superflow becomes
supercritical,
i.e. the superfluid velocity exceeds the Landau critical velocity. So it is
necessary to extend the consideration of the Hawking type radiation to higher
energies, where the other mechanisms of the decay of the supercritical
superflow
can become important. For example, the radiated particles with energies outside
the "relativistic" region can be Andreev scattered back to the black hole. Thus
both partners (particle and hole) of the Hawking radiation will remain
within the
horizon. This means that the particle creation in high gravity field can
disturb
the Minkowski quantum vacuum inside the horizon without any radiation to the
exterior. In principle such pair creation inside the horizon can be more
important for the dissipation of the "superluminal" (supercritical) superflow
than the Hawking radiation.

\end{document}